# Controlled Electrokinetic Particle Manipulation using Paper-and-Pencil Microfluidics

Sankha Shuvra Das[a], Shantimoy Kar[b], Sayantan Dawn[c], Partha Saha[a], Suman Chakraborty[ab]*

[a]Department of Mechanical Engineering, Indian Institute of Technology Kharagpur, Kharagpur, India-721302

[b]Advanced Technology Development Centre, Indian Institute of Technology Kharagpur, Kharagpur, India-721302

[c]Department of Mechanical Engineering, Jadavpur University, Kolkata, India 700032

*email: suman@mech.iitkgp.ernet.in

## ABSTRACT

Dielectrophoresis is a very promising technique for particle manipulation on a chip. In this study, we demonstrate a controlled mannuvering of polystryene particles on a simple 'paper-and-pencil' based device by exploiting the underlying electrokinetics with primary contribution from dielectrophoretic (DEP) forces. On contrary to other reported DEP devices, the present configuration does not demand a shophitcated laboratory module for creating a non-uniform electric field, which is essential requirement in DEP settings. We demonstrate positive dielectrophoresis (pDEP) to trap 1 µm size polystyrene particle for low-conductivity suspending medium, at an applied field strength of ~100 V/cm. In addition, the switching of the trapping direction (positive to negative dielectrophoresis) can be simply achieved by manipulating the conductivity of the media. We further bring out an optimum range of pH for effective particle trapping. These results have significant implications towards designing cell-on-a-chip based point of care diagnostic devices for resource limited settings.

# INTRODUCTION

Aggregation or trapping of cells has paramount importance towards understanding of several biological processes relevant to disease detection and management.[1,2] To diagnose a disease, it is often required to concentrate the samples to obtain a detectable signal.[3] One common practice is to use electric fields (especially dielectrophoresis (DEP)[4]) for controlled manipulation of particles or cells/ biological entities. DEP force arises when a dielectric particle experiences a dipole moment under a non-uniform electric field, subjected to either AC or DC bias. DEP-based manipulation has emerged as a promising electrokinetic tool, and is widely been used in the field of biomedical for sorting or trapping of proteins,[5] DNA,[4] bacteria,[6] synthetic colloidal particles[7] etc.

Until now, most of the conventional DEP processes have utilized metal electrodes for generating inhomogeneous electric field gradient using low-amplitude AC electric field, widely known as electrode-based DEP (eDEP).[8,9] Despite being efficient, this methodology has several inherent limitations such as electrode fouling, electrolysis etc. which essentially restrict its utilitarian scope. In this context, insulator based DEP (iDEP), deploying an array of insulating structures within the channel for creating an electric field gradient, has emerged as an effective alternative.[10–13] The electrodes are placed at the inlet and outlet reservoirs of the channel (far away from the observation window) and hence the samples are not in direct contact with the electrodes. As the electric field is applied, flux lines bend around the patterned insulating structures, leading to a non-uniform electric field, which further assist in particle accumulation near high electric field locations. The iDEP systems are robust and cost-effective since these eradicate the scope for electrode fouling as well as the requirement of specific material deposition for electrode fabrication. Hence, the system can be used for continuous particle manipulation unlike to any eDEP based systems. Furthermore, the system eliminates the requirement of external pumping as the samples are transported continuously through the combined effect of electrokinetic forces (like induced electroosmotic, electrophoretic and dielectrophoretic).[14]

In DC-iDEP systems, if the applied potential is low, the electroosmotic and electrophoretic forces (i.e. linear electrokientc phenomenon) supersede the DEP effect, leading to streaming

DEP to occur. In contrast, if the applied potential is high enough, DEP force dominates over other electrokinetic forces, opening up the paradigm of non-linear electrokinetic phenomenon resulting in particle trapping.[15] However, the requirement of higher potential for creating electric field gradient in DC-iDEP systems causes Joule heating that may affect the viability of biological entities.[16,17] Thus, usage of DC-offset AC electric field may be an effective alternative to reduce Joule heating, where the requirement of electric potential for particle trapping is relatively less as compared to the conventional DC based iDEP systems.[10] On the contrary, use of reservoirs based dielectrophoresis (rDEP), where the electric field gradients are generated inherently at the reservoir-channel constriction at lower applied voltages,[18–21] may also eradicate Joule heating issue. The rDEP systems are used to concentrate particles at the reservoir-channel junction. However, they are limited due to channel clogging upon continuous use, and is mainly used for short-range analysis. Furthermore, the use of different electrode configurations such as concentric electrodes to create a non-uniform electric field is another efficient approach for manipulation of E-coli, DNA, quantum dots etc.[22–24]

So far, first generation microfluidic devices (i.e. devices made of glass, PMMA and PDMS)[13,25,26] have been extensively used for particle trapping. However, the fabrication of these devices necessitates a sophisticated laboratory control and skilled personnel. This constraint demands for a simple and frugal device for particle manipulation. In last decade, paper-based microfluidics and thread-based microfluidics received substantial popularity due to simple and easy fabrication and can be easily operable from resource-limited places.[27–30] The liquid imbibes through the porous network of the fiber due to the capillary action. In order to improve the imbibition rate further, an electric field can be applied across paper-based devices through the graphite-sketched electrode.[31] Moreover, the inherent functionality of the device makes it useful for various applied and fundamental studies such as disease detection,[32] water quality monitoring,[33] power generation,[34] electrowetting,[35] micromixing[36] etc.

Here, we report the use of a 'paper-and-pencil' based platform to exploit DEP for controlled entrapment of the particles. We have used an innovative approach of concentrating polystyrene particles on paper-based devices using an asymmetric design of graphite electrodes (a pair of circular and rectangular shape electrode), which further affirms the non-uniformity in the applied electric field. We show that the particle trapping is possible by applying a low-magnitude DC electric field (~100 V/cm). Furthermore, the switching of trapping direction (positive DEP to negative DEP trap) is possible by manipulating the conductivity of the suspending medium. The particle immobilization obtained through the present method is irreversible in nature and thus may be helpful in various bio-medical detection processes where significant signal amplification is on demand.[37] The device can be further used as an effective alternative of any eDEP based devices without any scope of electrode fouling.

**THEORETICAL BACKGROUND**

When a particle is subjected to a non-uniform electric field, it experiences a net dielectric force that causes the particle to move toward the higher or lower electric field intensity regions. When the particle moves toward the higher electric field regions, the process is known as positive dielectrophoresis (pDEP). On the other hand, if the particle moves toward the lower electric field regions, it is termed as negative dielectrophoresis (nDEP). In case of DC electric field, the net dielectric force acting on a particle is given as,

$$F_{DEP} = 2\pi R_p^3 \varepsilon_m f_{CM} \nabla E_{DC}^2 \qquad (1)$$

where, $R_p$ is radius of the particle, $\varepsilon_m$ is permittivity of the medium, $E_{DC}$ is the applied DC electric field, $f_{CM}$ is the Clausius-Mossotti factor which can be defined as,

$$f_{CM} = \frac{\sigma_p - \sigma_m}{\sigma_p + 2\sigma_m} \qquad (2)$$

Here, $\sigma_p$ and $\sigma_m$ is the conductivity of particle and medium respectively.

In addition to DEP force, the particle also experiences linear electrokinetic forces, as governed by

$$V_{EK} = \mu_{EK} E_{DC} \qquad (3)$$

where, $\mu_{EK}$ is the linear electrokinetic mobility of the particle. In order to achieve DEP trapping, the electric field gradient is required to be sufficiently high so that the particle velocity due to DEP force supersedes the same due to linear electrokinetic force. Under such circumstances, the DEP velocity can be expressed as,

$$V_{DEP} = -\mu_{DEP} E_{DC}^2 \qquad (4)$$

where, $\mu_{DEP}$ is DEP mobility.

**EXPERIMENTAL DETAILS**

The device is fabricated on a standard laboratory grade Whatmann filter paper (grade 1, mean pore size ~11µm), where it is cut to the required design using a scissor. The device consists of two segments, a wider section (width ~12 mm and length ~25 mm) is connected to a thin section (width ~ 3 mm and length ~25 mm). In order to create a non-uniform electric field, we use asymmetric design of electrodes, which comprises a semi-circular electrode with inner and outer diameter of ~6 mm and ~12 mm, and a rectangular shape electrode with dimensions ~3 × 8 mm. The electrodes are sketched using 2B graphite pencil keeping the inter electrode spacing of ~2 mm. Furthermore, to make electrical connections, copper wires of ~350 µm size are connected to the graphite electrodes using conductive silver paste (Alfa Aesar). To apply DC electric field, an external source-meter (Keithley 2410) is connected to the copper electrodes.

The schematic of the experimental setup including the digital image of actual 'paper-and-pencil' based DEP device used for particle trapping is shown in Figure 1a and 1b. Experiments are performed at different applied DC electrical potentials: 5 V, 10 V and 20 V. Table 1 shows the different types of suspending media and their properties (electrical conductivity and pH) used in this study. Here, we use different concentrations of KCl solutions to investigate the trapping efficiency. The solutions are seeded with fluorescent particles (carboxylate modified polystyrene particles) of 1 µm size (FluoSpheres) with concentration of ~ 0.12 % (v/v). In order to compensate the evaporative loss and to avoid drying, the device is connected to an external reservoir, containing electrolyte solution, through a thin paper strip of ~ 2 mm width (refer to Figure 1b). A over-hanged segment of the device helps in creating a continuous flow of the liquid. Before dispensing the particle laden-electrolyte solution, it is ensured that the paper surface is completely wet and saturated with a thin layer of same solution. A constant volume of ~ 20µl sample solution is pipetted into the IEG.

During the experimentation, we capture time-lapse fluorescence images after a fixed interval of time (~ 1 min) over the maximum time period of 8 minutes (after which no further particle movement is observed) using an inverted fluorescence microscope (Olympus-IX71). In order to investigate the trapping efficacy quantitatively, mean grayscale intensity distribution (essentially represents the concentration profile) of the particle is measured along the inter-electrode gap (i.e. along the axial direction) of the device.

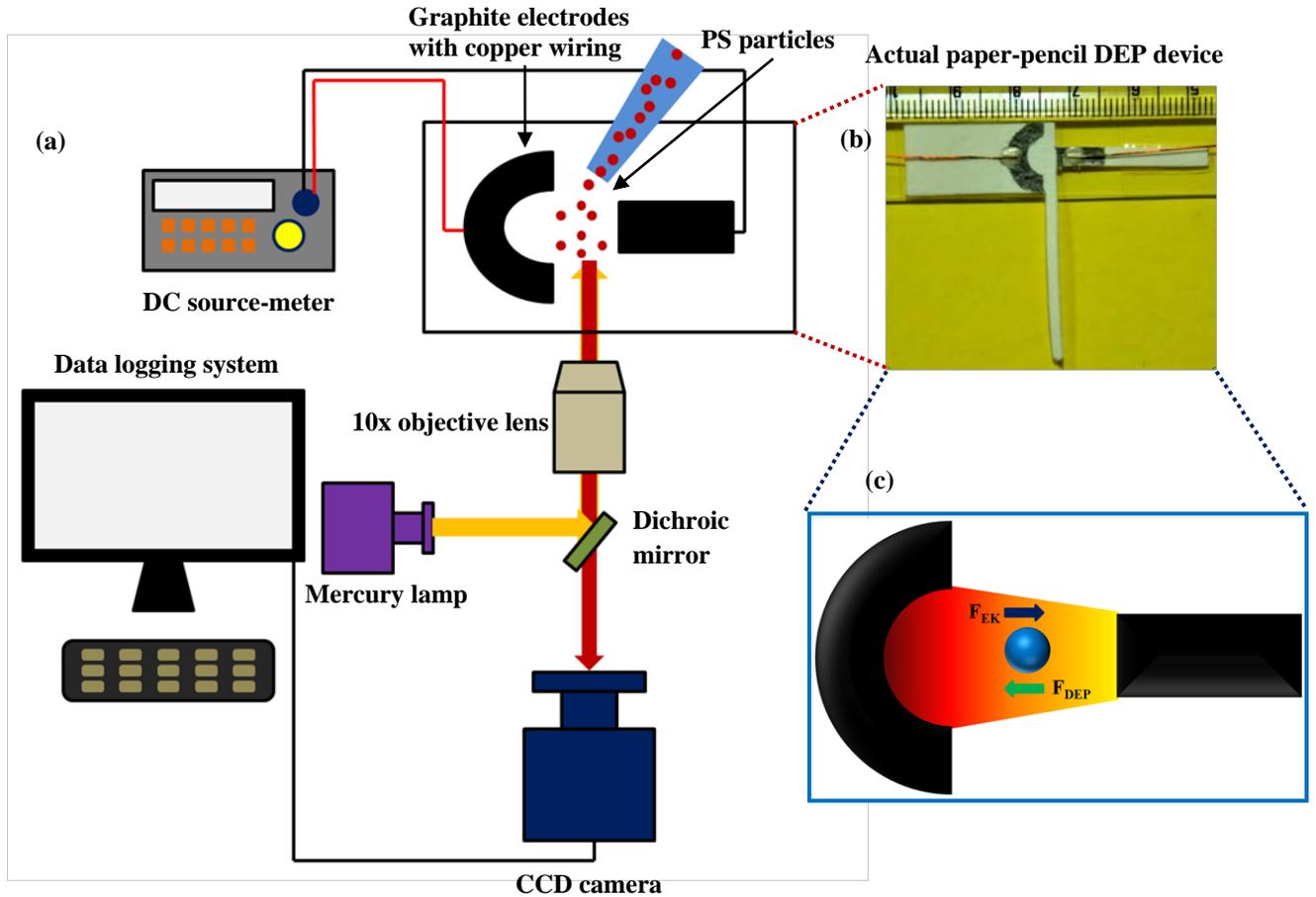

**Figure 1.** (a) Schematic representation of the experimental setup used for trapping of fluorescent-tagged charged polystyrene particles using DC electric field, (b) pictorial image of 'paper-and-pencil' DEP device, (c) schematic representation of different forces acting on a single particle, subjected to a DC electric field. Circular graphite-electrode connected to the positive and rectangular graphite-electrode connected to the negative terminal of the DC power source.

The intensity profile represented here is the y-averaged intensity along x-direction. The intensity at any specific x-location can be calculated as, $I(x) = \sum I_y / k_y$, where $k_y$ is the number of pixels along y-direction of a particular frame. The mean intensity is further normalized as, $I(x)_n = \sum I(x) / I_0$, where $I_0$, the highest possible intensity (i.e. 255). Towards the end (i.e. after completion of the experiment), we capture series of fluorescent images of the paper-matrix along the IEG of the channel (i.e. from negative to positive electrode), ensuring the overlapping of two consecutive frames.

Table 1 Type of electrolyte solutions and their properties used for the experimentation.

| Type of electrolyte | Concentration | Conductivity (µS/cm) | pH |
|---|---|---|---|
| DI water | 0 mM | 1.2 | 6.7 |
| KCl | 1 mM | 140 | 5.5 |
| KCl | 5 mM | 615 | 5.3 |
| KCl | 10 mM | 1165 | 5.2 |
| PVP (mixed with 1mM KCl) | 1% (w/v) | 170 | 4.1 |
| KOH | 1 mM | 360 | 10.6 |
| HCl | 1 mM | 220 | 3.1 |

## RESULTS AND DISCUSSIONS

When the cellulose fibres of the paper device come in contact with particle-laden electrolyte solution (1mM KCl), an electrical double layer (EDL) is formed near the fibre surface with negatively free surface charge (free carboxylic acid and hydroxyl group).[34] Thus, if the electric field is applied, particles in the suspending medium experience different forces viz. linear electrokinetic force and DEP force. For relatively larger particle size (~1 µm size) and low surface charge density of the particle (~20 mC/m$^2$), electrophoretic (EP) force can be neglected and linear electrokinetic (EK) force is thus assumed to be equivalent of the electroosmotic (EO) force.[38] For a given applied electric field, EO force strongly depends on the bulk liquid properties and the surface property of the device material. Thus, EO force acts towards the direction of an applied electric field and drives the bulk liquid (along with the particle) towards the negative electrode. On the other hand, the direction and the magnitude of the DEP force depends on several factors like the conductivity difference between particle and the suspending medium, electrode spacing, geometry of the device etc. The experimentally measured conductivity of the 1mM KCl solution ($\sigma \sim 140$ µS/cm) is higher than that of the carboxylate modified polystyrene particle ($\sigma \sim 40$µS/cm)[39] and thus the estimated trapping factor is, $f_{CM} \sim -0.31$ (using Eq. 2). Hence, according to the conventional DEP theory, particles would experience a negative DEP (nDEP) force and move towards the lower electric field region. However, we observe a reverse phenomenon for 1mM KCl solution (Figure. 3 and 4); where the particles move towards the high electric field region (positive electrode), reminiscent of the situation conforming to a positive DEP (pDEP) force. This observation may attribute to the polarization effect of the particles at DC applied electric field.[39] The control study (Figure S1) further reveals that no particle accumulation near either of the electrode is observed in absence of any electric field.

In general, carboxylate-modified polystyrene (PS) particle contains negative zeta potential which further depends on the conductivity and pH of the suspending medium. The experimentally measured zeta potential (Zetasizer; Malvern Instruments) of carboxylate modified PS particle suspended in DI water and 1mM KCl solution is ~ -42 mV and ~ -73mV respectively. Higher zeta potential corresponds to the larger EDL thickness with further enhancement in the surface conductance of the particle. Thus, the effective conductivity of the particle is assumed to be higher than that of the suspending medium and hence the particle exhibits a pDEP force and migrates towards the higher electric filed region. Trapping would only take place when the DEP force overpowers the EO force and hence particle trapping is observed near the positive electrode under the applied potentials. Figure 2 shows the temporal evolution of the particle concentration near positive electrode at 5 V DC applied potential (field strength of ~100 V/cm). The initial nucleation begins almost after ~ 2 min from the application of the electric field (Figure 2a). However, a significant trapping takes place at the end of ~ 5 min with complete migration of particles within ~ 8 min. The mean intensity distribution of the particle (Figure 2b) with concentration peak stems from the trapping of particles near the positive electrode. The peak concentration of the particles increases with time, with parabolic concentration profile, which signifies a maximum exerted pDEP force on particle at the centre of the region of interest. Analogous phenomenon is also observed for other applied potentials such as 10 V and 20 V with complete migration time of ~ 5 min and ~3 min, respectively.

Figure 3 shows the fluorescent microscopy images for different applied potentials, demonstrating the variation of particle concentration along the inter-electrode gap (IEG). The concentration profile (i.e. mean intensity) for 1mM KCl represents the trapping location (refer, Figure. 4a). We observe that, for an applied DC potential of ~5 V, the particles migrate at a very rapid rate and get trapped as well as form aggregates throughout the positive electrode surface. As the applied potential increases, DEP force increases with the square of the DC electric field (Eq. 1), and thus higher particle mobility is observed at 10 V and 20 V. It is further observed that the particles are primarily concentrated close to the silver pasting region, with several secondary trapping locations throughout the other sections of the circular electrode. The maximum trapping occurs at 5 V applied potential (refer, Figure. 4a), with relatively wide and sharp trapping band. Similar trapping efficacy is also observed at other potentials such as at 10 V, but with reasonably narrow trapping zone. When the applied potential exceeds further, the trapping efficiency reduces and comparatively less agglomeration is observed at 20 V, as compared to the other cases. In addition, current starts flowing at higher electric field which further enhances the possibility of Joule heating and thereby affects the particle concentration near the positive electrode.

Furthermore, to confirm the sole effect of DEP on particle trapping, we perform the same experiment with 1% (weight) polyvinypyrrolidone (PVP) mixed with 1mM KCl solution (see Table 1 for details). PVP prevents the formation of EDL near the fibre surface, and thus dampens the effect of EO force effecting the trapping.[40] The fluorescent microscopy images (Figure S2) and the mean intensity plot (Figure 4b) for different applied DC potentials show a significant particle concentration (equivalent to 1mM KCl) near the positive electrode. The highest concentration peak is achieved at 5V, with secondary peaks at different locations along the IEG. On contrary to 1mM KCl solution, particle accumulation only takes place near the positive electrode without any secondary trapping zone, which is probably due to the sole effect of DEP.

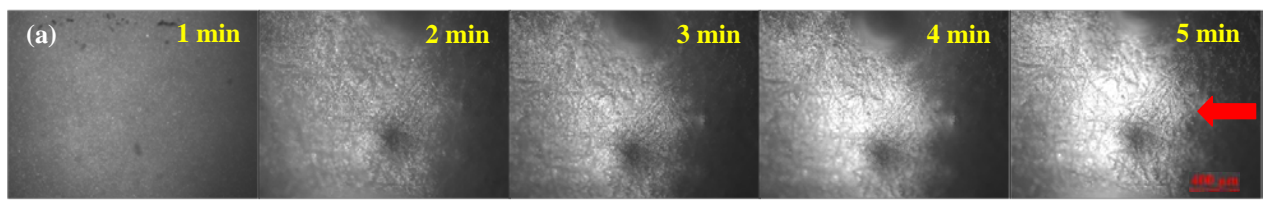

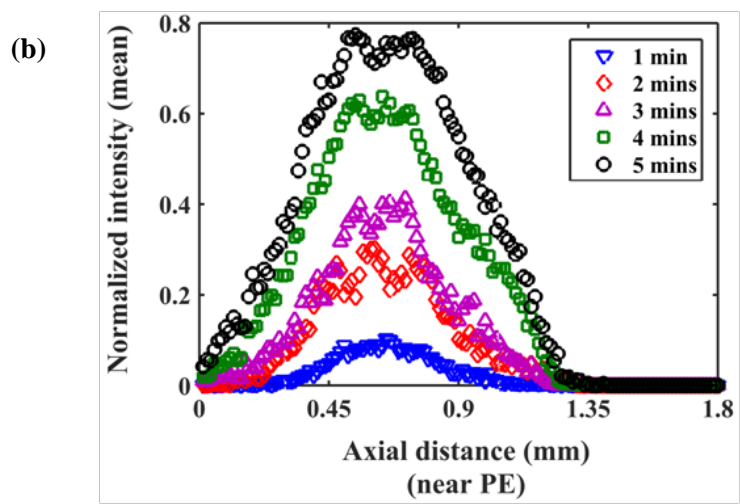

**Figure 2.** Time-lapse plot showing (a) particle aggregation, and (b) mean intensity variation of trapped particle near positive electrode with 1mM KCl suspending medium at 5V applied DC potential.

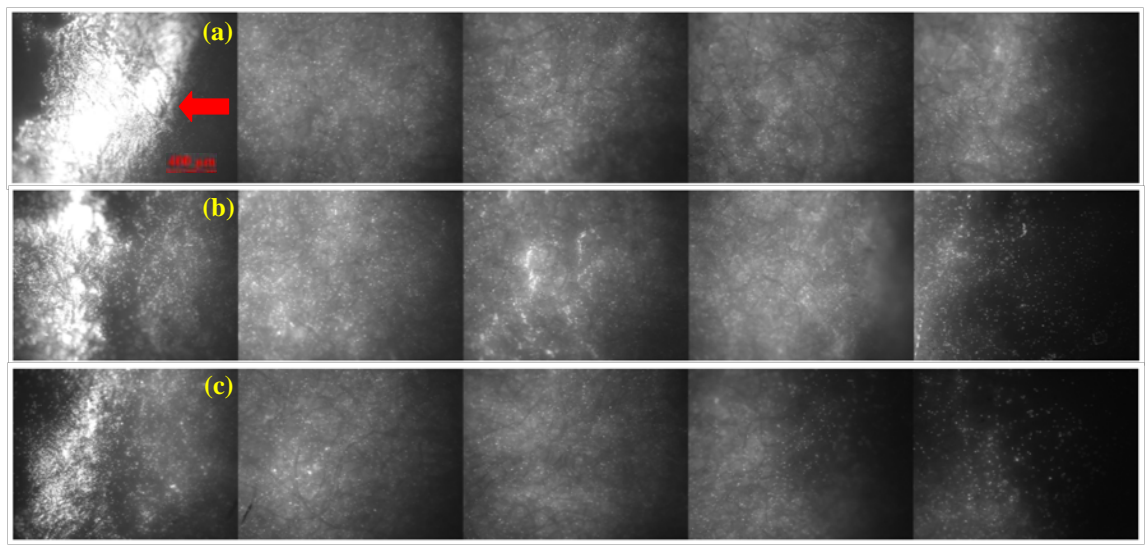

**Figure 3.** Fluorescence microscopy images depicting the distribution of trapped polystyrene particle at different locations along IEG (LHS: near positive electrode, RHS: near negative electrode) for 1mM KCl suspending medium at applied DC potentials of (a) 5V (b) 10V and (c) 20V. Red arrow indicates the particle trapped location at 5 V applied electric potential.

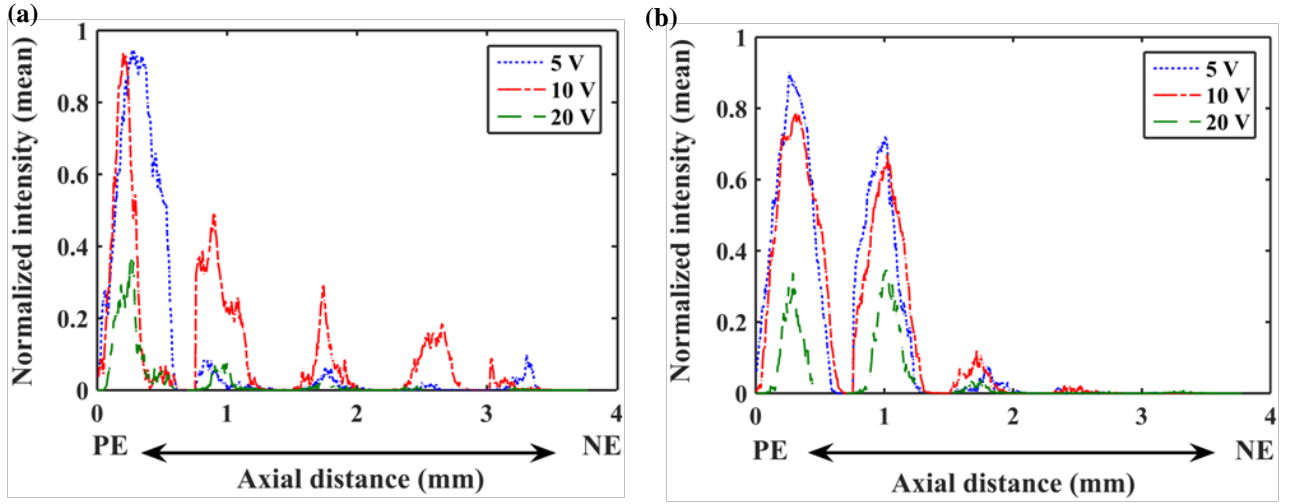

**Figure 4.** Spatial variation of mean intensity along the IEG for fluorescent-tagged polystyrene particle suspended in (a) 1mM KCl solution, (b) PVP mixed 1mM KCl solution (LHS: near positive electrode (PE), RHS: near negative electrode (NE) at different applied DC potentials.

From an earlier report[39], it is realized that low pH suspending medium favours for efficient particle trapping. In order to investigate the DEP characteristics with respect to different suspending medium properties (particularly with respect to conductivity and pH), we perform the experiment with 5mM ($\sigma \sim$ 615 µS/cm, pH ~5.3) and 10 mM KCl ($\sigma \sim$ 1165 µS/cm, pH ~5.25) solution at 5V applied potential. In contrary to previous observations, here particle migration is observed towards the lower electric field region. According to the electrokinetic theory, EDL thickness reduces as the medium conductivity increases. In such a context, the surface conductance of the particle is no longer a dominating factor, and hence the particle experiences a negative dielectrophoresis (nDEP) force (estimated as $f_{CM} \sim -0.31$) and migrates as well as gets trapped near the negative electrode. The mean intensity plot at 5V applied potential (refer, Figure 5) shows that at 5mM KCl solution ($\sigma \sim$ 615 µS/cm), significant particle trapping takes place at the mid-location of IEG along with negligible trapping near negative electrode. However, for 10 mM KCl solution, considerable trapping takes place near the negative electrode (see, Figure 5 and Figure S3). Nevertheless, the mean intensity obtained for 10mM KCl is less than that of 1mM KCl solution. Thus, a transition or switching from pDEP to nDEP is observed at 5 mM KCl solution. Such a switching phenomenon can also be useful for selective trapping of the particles in a particular location by manipulating the medium conductivity.

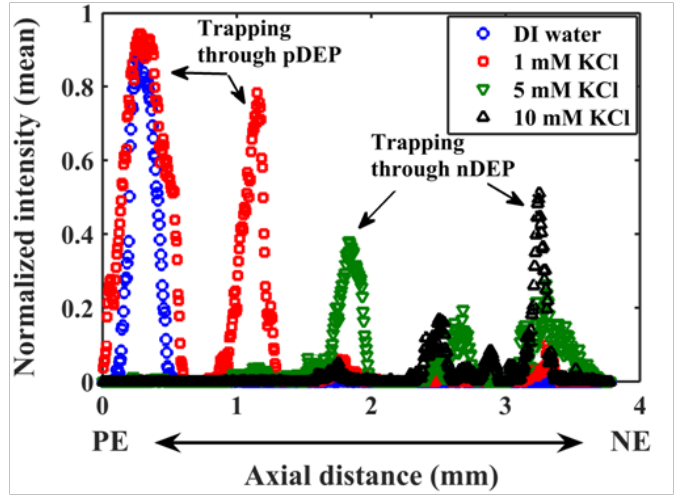

**Figure 5.** Normalized mean intensity variation of trapped particle along the inter electrode gap for suspending mediums with different conductivity. Switching from pDEP to nDEP trapping occurred in between 1mM to 5mM KCl solution.

We pursue further experiments with DI water to investigate the effect of particle conductivity on DEP trapping. As the conductivity of the suspending medium decreases, the effect of EK forces decreases and hence particle experiences DEP force only. The conductivity of DI water ($\sigma \sim 1.2$ µS/cm) is less than that of the PS particle, and hence the estimated trapping factor is $f_{CM} \sim 0.91$, which further signifies a high trapping efficiency. Under such circumstances, particle experiences a pDEP force and migrate towards the positive electrode. It is interesting to note that unlike the previous cases, here the particles are trapped only at the periphery of the silver pasting area (silver electrode, $E_s$) rather than other locations of the circular electrode (see, Figure S4). The particle concentrates to a narrow region due to effect of DEP only, and thus highly localized trapping (no secondary trapping location) is observed (refer, the concentration profile in Figure. 6). We further measure the elctrokinetic mobility of the particle through µPIV measurements. The particle average velocity, $V_p$, under different applied DC electric fields for different suspending media is shown in Figure 7. The average velocity varies between ~ 500-950 µm/s for particles suspended in DI water, whereas it is of ~ 235-290 µm/s for particles suspended in 1mM KCl solution at 5V applied potential. The existence of EO force in 1mM KCl acts opposite to the DEP effect and thereby reduces the electrokinetic mobility.

On the other hand, significant trapping can also be observed for higher conductivity suspending mediums. The thickness of EDL reduces as the conductivity of the suspending medium increases, and hence trapping is more significant due to sole effect of DEP. In this study as the IEG is small, we perform the experiment with lower conductivity solution to avoid Joule heating. This method of particle trapping is highly economical and more efficient than the reported strategies where microfluidic devices with cylindrical insulating structures are used for selective trapping of particles at relatively higher voltages (~750 V DC).[38] In order to reduce the requirement of such higher voltages, 3D microfluidic devices with insulating structures are used.[25] The 3D-iDEP device can effectively trap particles at 10 V applied potential, which is significantly higher than that for our case, where significant trapping is observed at 5 V. Our simple 'paper-and-pencil' based DEP device can efficiently trap particles without requiring any sophisticated laboratory control which further eliminates the overall operational costs.

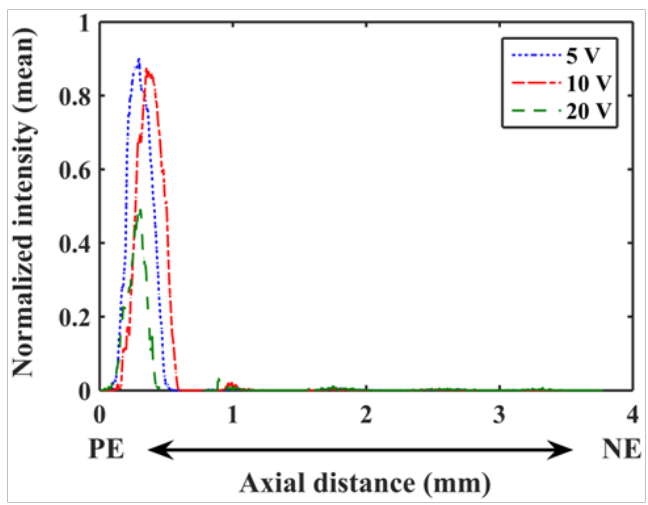

**Figure 6.** Spatial variation of mean intensity along the IEG for fluorescent-tagged polystyrene particle suspended in DI water (LHS: near positive electrode, RHS: near negative electrode) at different applied DC potentials.

To further investigate the effect of pH on trapping efficiency, we perform the same experiment with different pH solutions such as 1mM KOH (pH ~10.6), HCl (pH~ 3.1) and then compare with the results corresponding to 1mM KCl solution. We observe that as the pH of the solution increases, particle concentration near positive electrode decreases (Figure 8). In general, as the pH of the solution increases, the number density of $OH^-$ group in the solution increases, which further increases the negative charge density (along with the carboxylic group) at the fiber surface. Thus the zeta potential of the paper surface increases, which in turn enhances the effect of EO force over DEP force. Hence, less trapping is observed for KOH solution as compared to

KCl solution (see, Figure S5). On the other hand, to investigate the trapping efficiency in highly acidic medium, we further perform the experiment with 1mM HCl solution. We observe that particles are loosely concentrated near the positive electrode. It is further observed that the intensity of the particle reduces, which may attribute to the chemical interaction of the particle with acidic medium. Therefore, an optimum pH range to achieve an effective particle trapping on paper-and-pencil device is ~ 5.5 to 6.7.

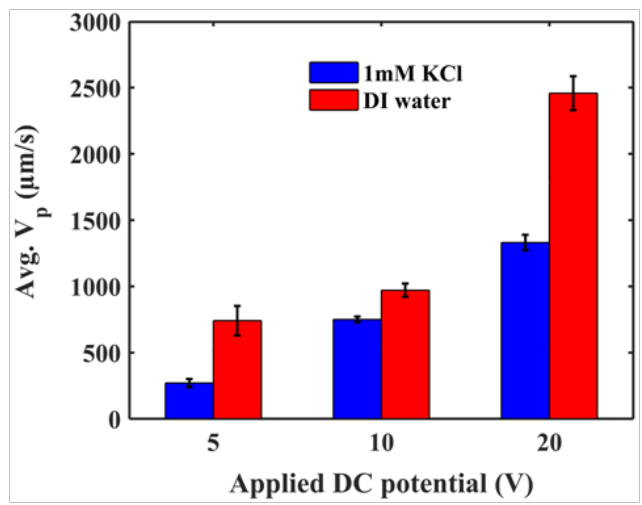

**Figure 7.** Average particle velocity suspended in different conductive medium against different applied field.

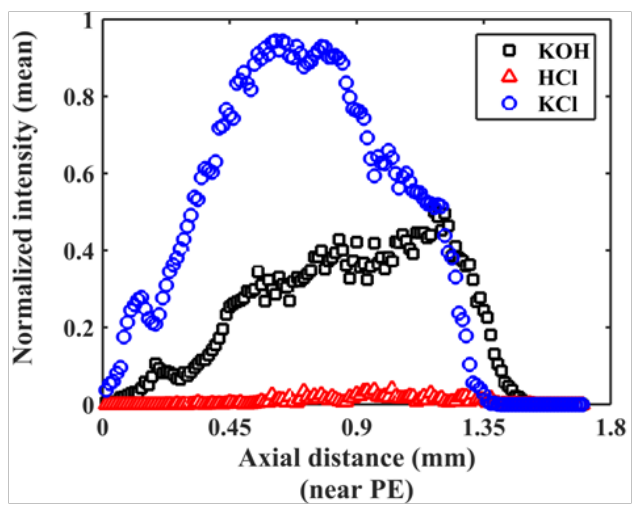

**Figure 8.** Normalized mean intensity near positive electrode, PE, for fluorescent-tagged polystyrene particle suspended in different pH electrolytes (Left side: near positive electrode, Right side: near negative electrode) at different applied DC potentials.

## CONCLUSIONS

In summary, a simple and frugal 'paper-and-pencil' based dielectrophoretic device is introduced for trapping of polystyrene particles. The trapping characteristics of negatively charged polystyrene particles of 1µm size are investigated under different DC applied potentials on paper-based platform. The present study unveils the dependence of suspending medium conductivity on the trapping location. The device primarily uses pDEP to trap particles near positive electrode; however exploits nDEP if the concentration of KCl solution is higher than 5mM. The linear elctrokinetic mobility of the particles confirm that particles suspended in DI water experience pure DEP force and hence the average particle velocity is higher than that corresponding to the scenario with1mM KCl (for which the particles experiences combined effect of linear electrokinetics and DEP force). Thus, in case of DI water, a highly localized trapping is observed near positive electrode. We further show that the effective particle trapping takes place within the pH range of ~ 5.5 - 6.7. We envisage that the present investigation lays the foundation of being considered to constitute the fundamental premises of designing pre-concentrator technique for different bio-analytical signal detection systems in low-cost paradigm.

## ACKNOWLEDGEMENTS

The authors would like to acknowledge Sponsored Research and Industrial Consultancy (SRIC) Cell, IIT Kharagpur for the financial support of the 'Plant on-a-Chip' project provided through the SGDRI grant.

## REFERENCES

(1) Perez-Gonzalez, H. V.; Gallo-Villanueva, C. R.; Cardenas-Benitez, B.; Martinez-Chapa, O. S.; Lapizco-Encinas, H. B. *Anal. Chem.* **2018**, *90*, 4310−4315.

(2) Hunt, T. P.; Westervelt, R. M. *Lab Chip* **2008**, *8*, 81-87.

(3) Shafiee, H.; Caldwell, J. L.; Sano, M. B.; Davalos, R. V. *Biomedical Microdevices* **2009**, *11*, 997–1006.

(4) Mahshid, S.; Lu, J.; Abidi, A. A.; Sladek, R.; Reisner, W. W.; Ahamed, M. J. *Sci. Rep.* **2018**, *8*, 1–12.

(5) Rito-palomares, M.; Lapizco-encinas, B. H.; Ozuna-chac, S. *J. Chromatogr. A* **2008**,

*1206*, 45–51.

(6) Park, S.; Zhang, Y.; Wang, T.; Yang, S. *Lab Chip* **2011**, *11*, 2893–2900.

(7) Ramos, A.; Morgan, H; Green, G., N.; Castellanos, A. *Journal of Colloid and Interface Science* **1999**, *217*, 420–422.

(8) Kang, Y.; Li, D.; Kalams, S. A.; Eid, E. J. *Biomed Microdevices* **2008**, *10*, 243–249.

(9) Regtmeier, J.; Eichhorn, R.; Viefhues, M.; Bogunovic, L.; Anselmetti, D. *Electrophoresis* **2011**, *32*, 2253–2273.

(10) Lewpiriyawong, N.; Yang, C.; Lam, C.,Y. *Microfluid. Nanofluid.* **2012**, *12*, 723–733.

(11) Srivastava, K., S.; Artemiou, A.; Minerick, A. R. *Electrophoresis* **2011**, *32*, 2530–2540.

(12) Lapizco-encinas, B. H.; Simmons, B. A.; Cummings, E. B.; Fintschenko, Y. *Electrophoresis* **2004**, *25*, 1695–1704.

(13) Ivory, C. F.; Srivastava, S. K. *Electrophoresis* **2011**, *32*, 2323–2330.

(14) Dey, R.; Shaik, V. A.; Chakraborty, D.; Ghosal, S.; Chakraborty, S. *Langmuir* **2015,** *31***,** 5952–5961.

(15) Cummings, E. B.; Singh, A. K. *Anal. Chem.* **2003**, *75 (18)*, 4724–4731.

(16) Xuan, X. *Electrophoresis* **2008**, *298*, 33–43.

(17) Gallo-villanueva, R. C.; Sano, M. B.; Lapizco-encinas, B. H.; Davalos, R. V; *Electrophoresis* **2014**, *35*, 352–361.

(18) Church, C.; Zhu, J.; Huang, G.; Tzeng, T.; Xuan, X. *Biomicrofluidics* **2010**, *4*, 044101.

(19) Patel, S.; Showers, D.; Vedantam, P.; Tzeng, T.; Qian, S.; Xuan, X. *Biomicrofluidics* **2012**, *6*, 34102.

(20) Zhu, J.; Hu, G.; Xuan, X. *Electrophoresis* **2012**, *33*, 916–922.

(21) Harrison, H.; Lu, X.; Patel, S.; Thomas, C.; Todd, A.; Johnson, M.; Raval, Y.; Tzeng, T.; Song, Y.; Wang, J.; Li, D.; Xuan, X. *Analyst* **2015**, *140*, 2869–2875.

(22) Wong, P. K.; Chen, C.; Wang, T.; Ho, C. *Anal. Chem.* **2004**, *76*, 6908–6914.

(23) Bown, M. R.; Meinhart, C. D. *Microfluid. Nanofluid.* **2006**, *2*, 513–523.

(24) Sin, M. L. Y.; Gau, V.; Liao, J. C.; Haake, D. A.; Wong, P. K. *J. Phys. Chem. C* **2009**, *113*, 6561–6565.

(25) Braff, W. A.; Pignier, A.; Buie, C. R. *Lab Chip* **2012**, *12*, 1327–1331.

(26) Li, Y.; Dalton, C.; Crabtree, H. J.; Kaler, K. V. I. S. *Lab Chip* **2007**, *7*, 239–248.

(27) Nie, Z.; Nijhuis, C. A.; Gong, J.; Chen, X.; Kumachev, A.; Martinez, A. W.;

Narovlyansky, M.; Whitesides, G. M. *Lab Chip* **2010**, *10*, 477–483.

(28) Ballerini, D. R.; Li, X.; Shen, W. *Biomicrofluidics* **2011**, *5*, *14105*.

(29) Hwang, H.; Kim, S.; Kim, T.; Park, J.; Cho, Y. *Lab Chip* **2011**, *11*, 3404–3406.

(30) Wang, X.; Hagen, J. A.; Papautsky, I.; Wang, X.; Hagen, J. A.; Papautsky, I. *Biomicrofluidics* **2013**, *7*, *14107*.

(31) Mandal, P.; Dey, R.; Chakraborty, S. *Lab Chip* **2012**, *12*, 4026–4028.

(32) Hu, J.; Wang, S. Q.; Wang, L.; Li, F.; Pingguan-Murphy, B.; Lu, T. J.; Xu, F. *Biosens. Bioelectron.* **2014**, *54*, 585–597.

(33) Loh, L. J.; Bandara, G. C.; Weber, G. L.; Remcho, V. T. *Analyst* **2015**, *140*, 5501–5507.

(34) Das, S. S.; Kar, S.; Anwar, T.; Saha, P.; Chakraborty, S. *Lab Chip* **2018,** *18*, 1560-1568.

(35) Kim, D. Y.; Steckl, A. J. *Applied Materials and Interfaces* **2010**, *2 (11)*, 3318-3323.

(36) Dey, R.; Kar, S.; Joshi, S.; Maiti, K. T.; Chakraborty, S. *Microfluid. Nanofluid.* **2015**, *19*, 375–383.

(37) Fu, E.; Kauffman, P.; Lutz, B.; Yager, P. *Sensors Actuators B. Chem.* **2010**, *149* (1), 325–328.

(38) Baylon-Cardiel, L. J.; Lapizco-Encinas, H., B.; Reyes-Betanzo, C.; Chavez-Santoscoy, V. A.; Martínez-chapa, S. *Lab Chip* **2009**, *197, 2896-2901*.

(39) Martínez-lópez, J. I.; Moncada-hernández, H.; Rito-palomares, M.; Lapizco-encinas, B. H. Anal. Bioanal. Chem. **2009**, *394*, 293–302.

(40) Milanova, D.; Chambers, R. D.; Bahga, S. S.; Santiago, J. G. *Electrophoresis* **2012**, *33*, 3259–3262.

# Electronic Supplementary Information

## Controlled Electrokinetic Particle Manipulations on 'Paper-and-Pencil' Devices

Sankha Shuvra Das[a], Shantimoy Kar[b], Sayantan Dawn[c], Partha Saha[a], Suman Chakraborty[ab]*

[a]Department of Mechanical Engineering, Indian Institute of Technology Kharagpur, Kharagpur, India-721302

[b]Advanced Technology Development Centre, Indian Institute of Technology Kharagpur, Kharagpur, India-721302

[c]Department of Mechanical Engineering, Jadavpur University, Kolkata, India 700032

*email: suman@mech.iitkgp.ernet.in

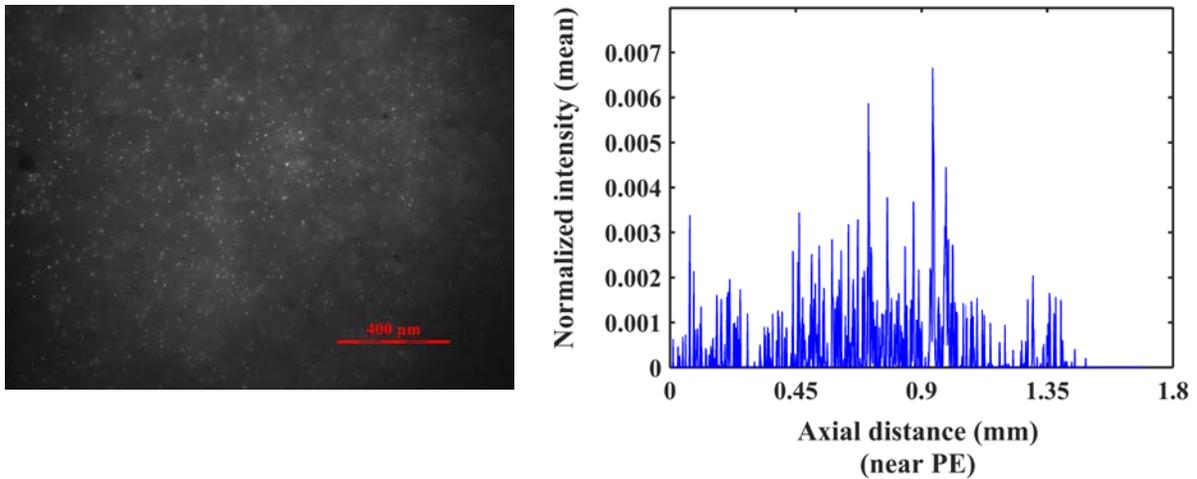

**Figure S1.** Particle distribution without any electric field (a control study), (a) fluorescence microscopy image and (b) normalized particle mean intensity indicates no such trapping of particles near positive electrode (PE).

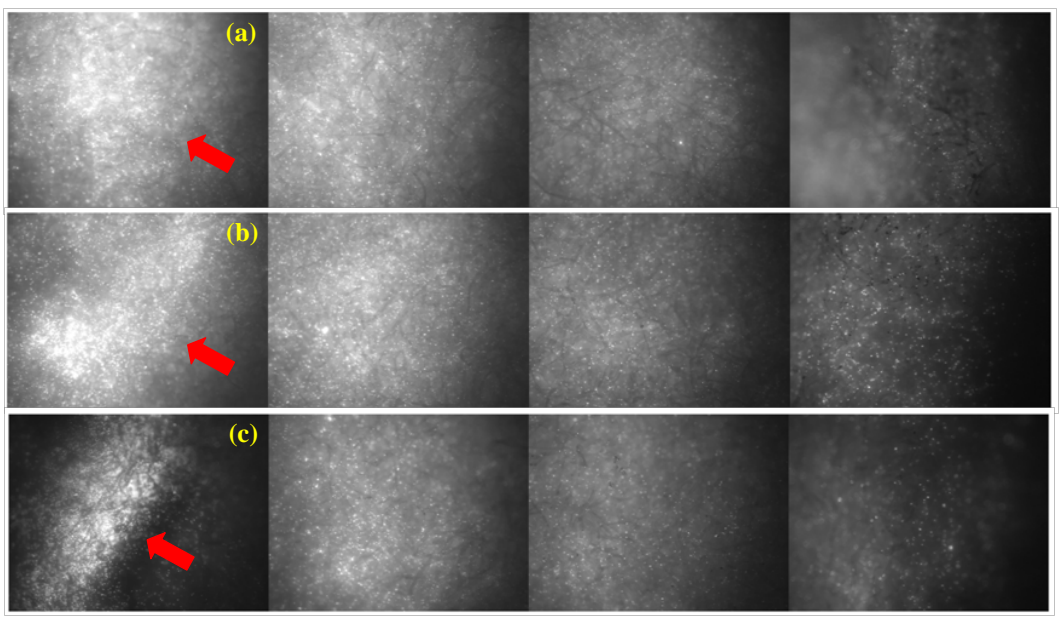

**Figure S2.** Fluorescence microscopy images depicting the distribution of trapped particle at different locations (LHS: near positive electrode, RHS: near negative electrode) for PVP mixed 1mM KCl at (a) 5V (b) 10V and (c) 20V.

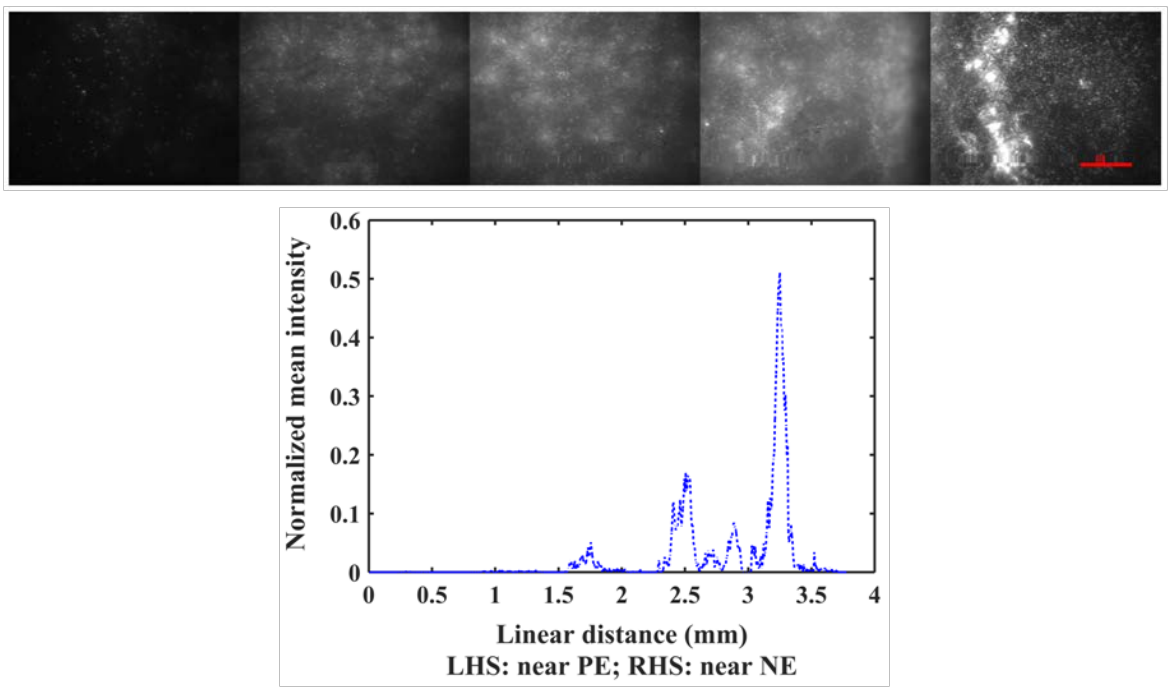

**Figure S3.** Spatial variation of particle concentration and the associated mean intensity along the IEG for fluorescent-tagged polystyrene particle suspended in 10mM KCl solution at 5V DC potentials. LHS and RHS represents the near positive and negative electrode side respectively for both the Figures.

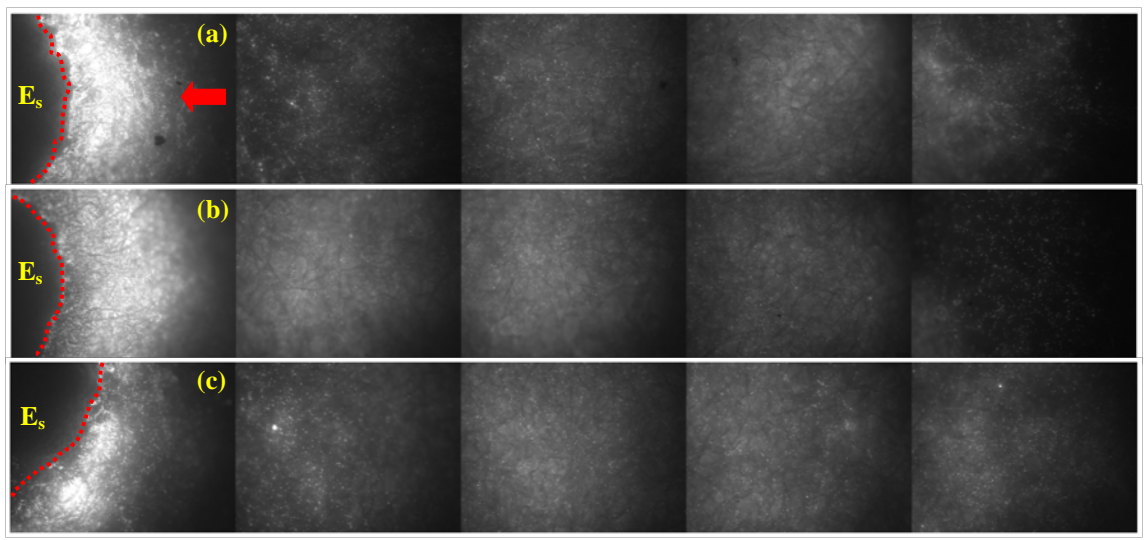

**Figure S4.** Fluorescence microscopy images depicting the distribution of trapped particle at different locations (LHS: near positive electrode, RHS: near negative electrode) for DI water at (a) 5 V (b) 10 V and (c) 20 V. Highly localized trapping is obtained in the vicinity of the silver pasting ($E_s$ indicates the area covered by silver electrode or silver pasting site; arrow indicates the trapped location).

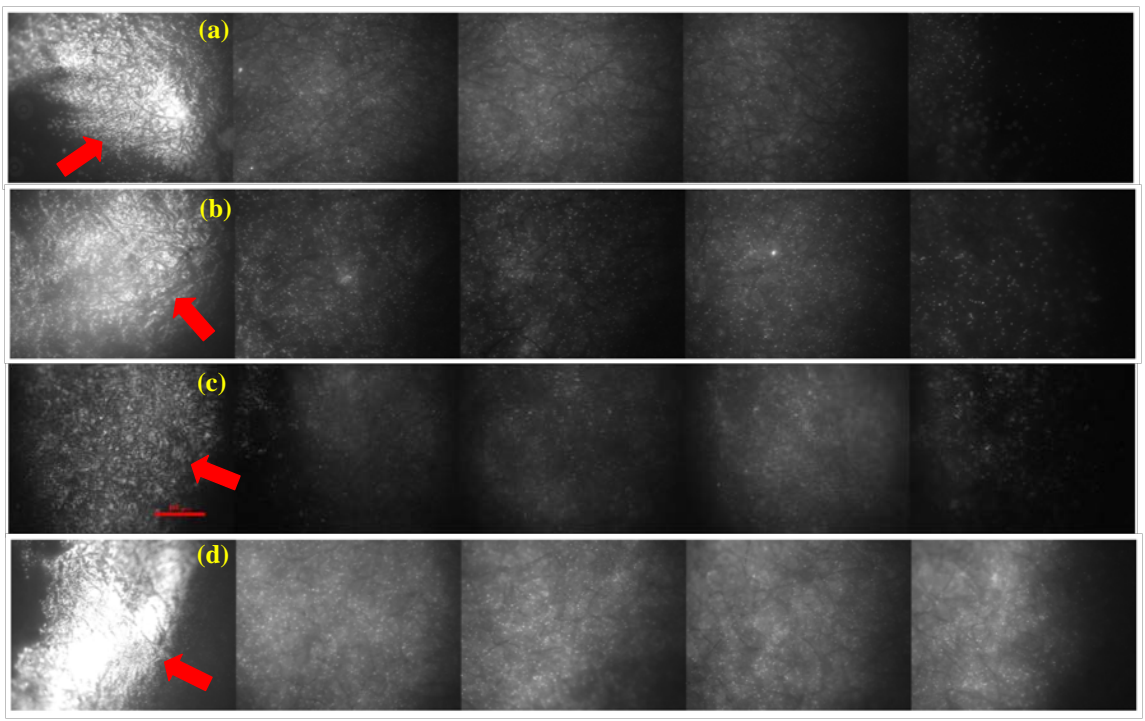

**Figure S5.** Fluorescence microscopy images depicting the distribution of trapped particle at different locations (LHS: near positive electrode, RHS: near negative electrode) for different pH solutions at 5V applied potential (a) 1mM KOH (b) 1mM NaOH (c) 1mM HCl (d) 1mM KCl.